\documentclass[10pt,conference,letterpaper]{IEEEtran}

\usepackage{cite}
\usepackage{amsmath,amssymb}
\usepackage{graphicx}
\usepackage{booktabs}
\usepackage{multirow}
\usepackage{multicol}
\usepackage{array}

\IEEEoverridecommandlockouts
\begin{document}

\title{High-Performance Low-Power Adiabatic Systolic Array Design in Advanced FinFET Nodes}


\author{
\IEEEauthorblockN{Jun Yin, Liangtao Dai, Yimin Gao, and Mircea R. Stan}
\IEEEauthorblockA{
Department of Electrical and Computer Engineering, University of Virginia\\
Charlottesville, VA, USA\\
\{htf6ry, kvf4sf, yg9bq, mircea\}@virginia.edu
}
\thanks{This research was supported in part by an endowed Virginia
Microelectronics Consortium (VMEC) professorship and by gift awards
from Intel, AMD, and Northrop Grumman.}
}

\maketitle

\begin{abstract}
Adiabatic logic has traditionally been recognized as a low-power solution but constrained to low clock speeds to preserve adiabatic behavior.
For advanced FinFET nodes, however, clock frequencies have plateaued due to power/thermal concerns (dark silicon) even as the {\em intrinsic device speeds have continued to scale}. 
This convergence opens an opportunity for adiabatic logic to maintain adiabatic behavior even at GHz clocks.
We demonstrate an adiabatic logic (AL) design methodology through a MAC systolic array 
implemented in commercial 16 nm FinFET technology with a resonant 4-phase power clock (PCK) generator, including digital-to-AL and AL-to-digital interfaces. 
Simulations show that the AL MAC systolic array at 1 GHz achieves power reductions of up to 42\% and 36\% at the core and system levels, respectively, compared to digital counterparts. 
Scaling to more advanced nodes should provide even better power/performance metrics.
\end{abstract}

\begin{IEEEkeywords}
Adiabatic logic, Systolic array, Multiplication-accumulation, 4-phase power clock generator, IoT, FinFET technology node, Low power
\end{IEEEkeywords}

\section{INTRODUCTION}
The 
proliferation of AI 
workloads has 
intensified demand for efficient hardware architectures capable of accelerating 
matrix multiplication, a fundamental operation across modern computing, image processing, and signal processing applications~\cite{wang2019flexible,jouppi2017datacenter,zhuang2023high}.
As model sizes and throughput requirements continue to grow, achieving high energy efficiency at sustained clock frequencies has become a central design challenge. 

Systolic arrays arrange multiply-accumulate (MAC) units in a regular two-dimensional grid, streaming data through orchestrated patterns 
that maximize data reuse and minimize memory bandwidth requirements~\cite{kung1979systolic,patil2018survey}. Their regular structure and high arithmetic intensity make them a preferred building block in modern AI accelerators, from Google's TPU \cite{jouppi2017datacenter} to numerous edge inference engines.
However, conventional digital systolic arrays face the $CV^2f$ power consumption curse
with power surging dramatically as frequencies increase \cite{yin2025unlocking}. 
Several architectural optimizations have been proposed
to improve 
systolic array efficiency. 
Approximate computing techniques \cite{waris2019design} reduce power but sacrifice accuracy, limiting applicability to error-tolerant workloads. 
Separating multipliers from adders in tree topologies \cite{asgari2020meissa} improves latency at the cost of increased area and hardware complexity. 
Dataflow optimizations such as the diagonal-input permutated weight-stationary (DiP) scheme \cite{abdelmaksoud2025dip} improve energy efficiency but still consume hundreds of milliwatts at 1 GHz, limiting suitability for edge deployment. These approaches all operate within the conventional digital $CV^2f$ power framework, leaving fundamental room for improvement.
\begin{figure}[!t]
\centering
\includegraphics[width=3.4in]{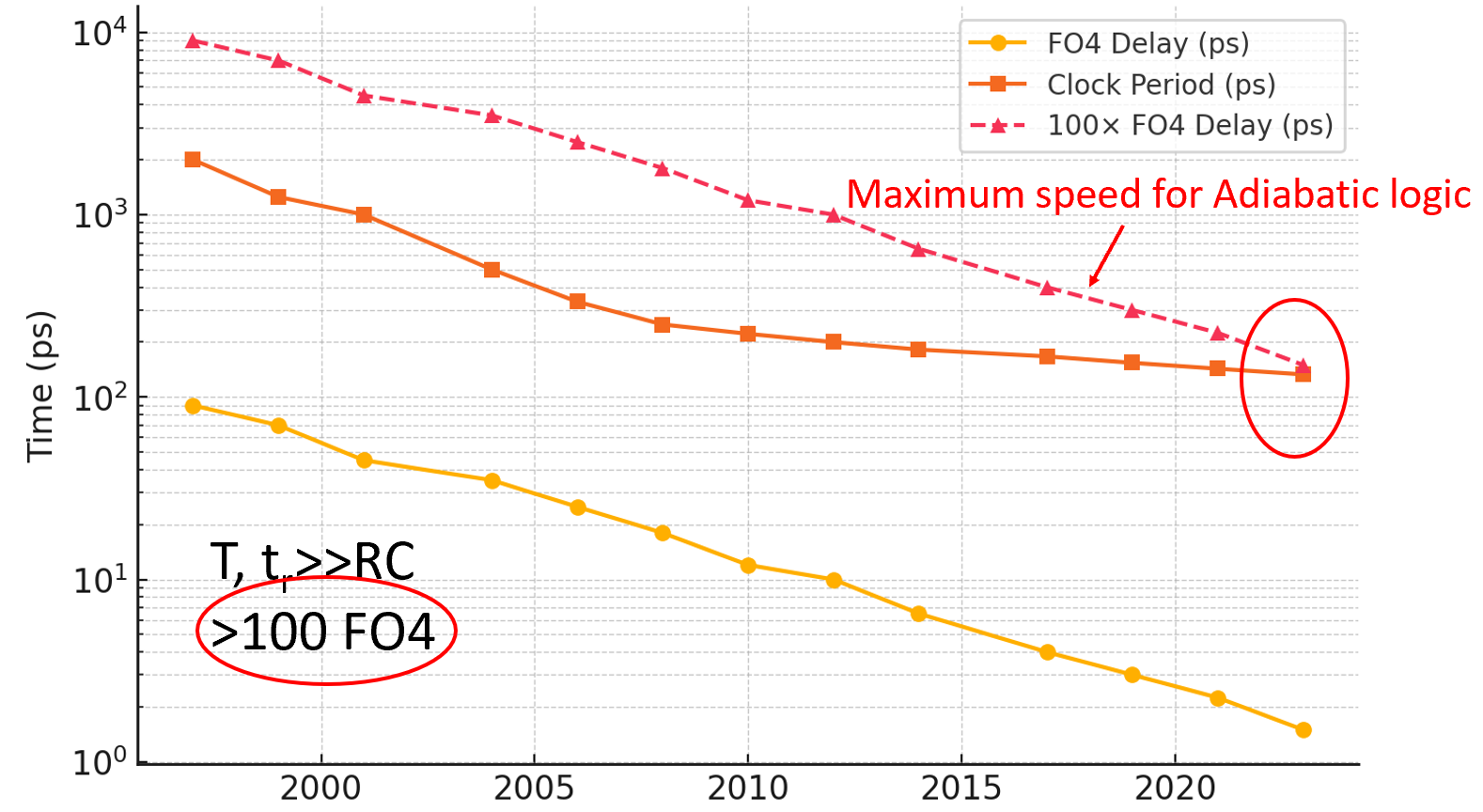}
\caption{Intrinsic FO4 delay, $100\times$ FO4 delay and clock period over the years. While the intrinsic FO4 delay has continued to scale, the clock frequencies/periods have plateaued. Traditionally adiabatic circuits have to run $100\times$ lower than their conventional counterparts in order to maintain adiabatic behavior, which, for advanced nodes can be obtained at about the same clock frequencies as conventional digital.}
\label{adiabatic_trends}
\end{figure}

Adiabatic Logic (AL) circuits~\cite{frank2021perfectly} offer a fundamentally different approach by recovering and reusing charge from load capacitances rather than dissipating it as heat. 
Historically, AL has been considered impractical for high-performance applications because maintaining adiabatic behavior requires transition times that far exceed the circuit's RC time-constant, forcing operation at low clock frequencies.
However, as shown in Fig. \ref{adiabatic_trends}, advanced FinFET technology nodes have driven RC time-constants into the picosecond range, while clock frequencies have plateaued at around 1 GHz due to power and thermal constraints. This convergence means that AL circuits can now maintain adiabatic behavior at the same clock frequencies used by conventional digital logic, opening the door to high-performance low-power accelerators for HPC and AI workloads.

We propose a design methodology for high performance low power AL through the circuit design of an AL MAC systolic array using positive feedback adiabatic logic (PFAL) gates on commercial 16 nm FinFET technology, with an efficient resonant 4-phase power clock (PCK) generator optimized to drive the array. The proposed AL circuits sustain 1 GHz clock rates with substantially reduced power, enabling a new generation of energy-efficient accelerators.
The major contributions are:
\begin{itemize}
    \item We introduce a bottom-up methodology to build PFAL gates, complex MAC blocks, and dual-rail digital to AL and AL to digital conversion circuits.
    \item We design a novel resonant 4-phase AC PCK sinewave generator using coupled 2-phase LC oscillators with shunt NFETs. To achieve high efficiency, the
    transistors are optimized
    such that the PCK generator can deliver sinusoidal clocks to AL circuits at low input power.
    \item
    We build and simulate a simple AL MAC systolic array integrated with digital FIFOs and converters and powered by the sinewave PCK.
    The AL MAC systolic array with peripherals and PCK included shows significant power reduction (low-power) while maintaining the frequency (high-performance) of conventional digital MAC systolic arrays.

\end{itemize}
\section{BACKGROUND}
\subsection{Traditional digital MAC}
\begin{figure}[!t]
\centering
\includegraphics[width=3.2in]{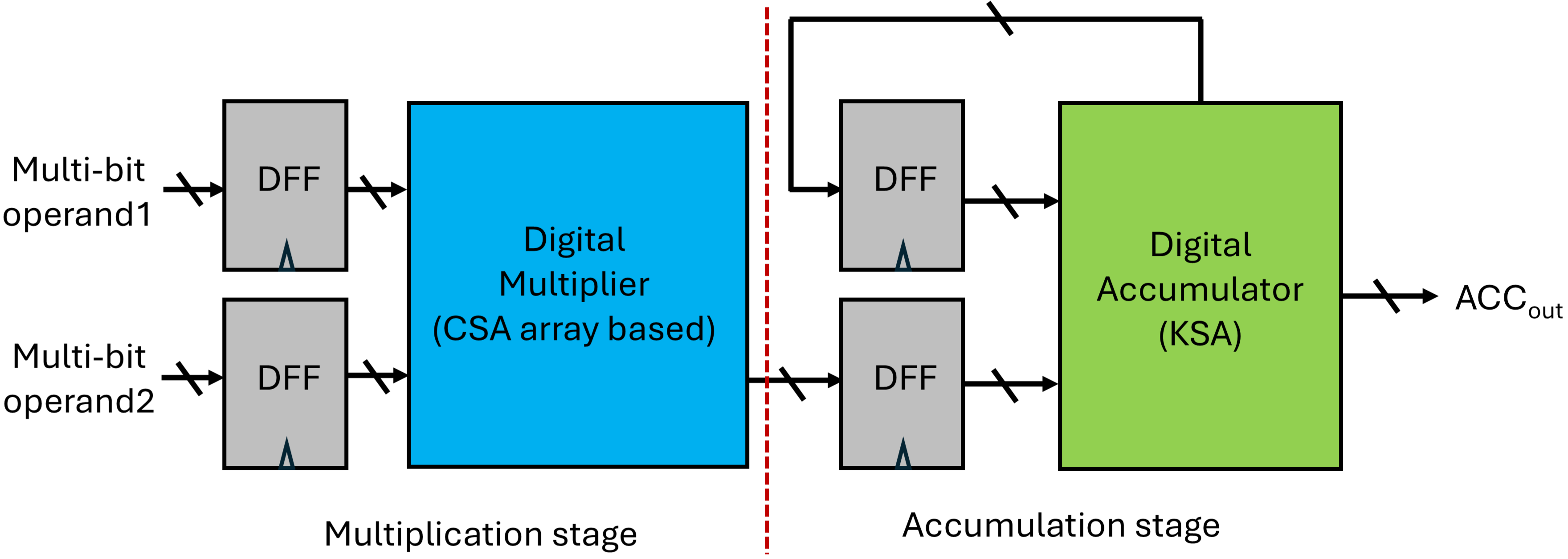}
\caption{Structural diagram of traditional MAC unit.}
\label{digital_mac}
\end{figure}
Fig. \ref{digital_mac} shows a digital MAC unit~\cite{cavanagh1983digital} that performs multiplication in the first stage and accumulation in the second. In multiplication, multi-bit operands are clocked through D-flip-flops (DFF) to carry-save adder (CSA)-based multiplier blocks generating partial products via bitwise AND operations. Parallel CSA arrays compress intermediate stages to two rows, avoiding carry propagation delays and enabling efficient shift-and-add operations, especially when pipelined. Intermediate registers balance delays between the multiplier and the accumulator stages for high clock frequencies. For accumulation, a Kogge-Stone prefix adder (KSA) propagates carries rapidly for parallel, efficient addition. On the first operation, the intermediate product is added with 0; on subsequent cycles, it is added with the previous accumulated sum latched in the accumulator register, which feeds back each cycle.
\subsection{Adiabatic logic with FinFET}\label{AL_funda}
AL is a design methodology that minimizes power
by recovering and reusing the charge stored on the transistor capacitances~\cite{samanta2009adiabatic}. Unlike traditional CMOS circuits that dissipate energy as heat, adiabatic logic uses a time-varying AC power supply to ensure transistors switch only when voltage differences are minimal and that the charging and discharging are slower than the time-constant of the circuit thereby reducing power loss. The energy
per cycle for adiabatic circuits ($E_\text{AL}$) is given quantitatively by \cite{wan2017energy}:
\begin{equation}\label{f1}
  E_\text{AL} = \frac{2CV_\text{dd}^2 RC}{t_\text{tran}}
\end{equation}

where $R$ is the ON-resistance of the transistor, $C$ is the capacitance of the circuits ($RC$ is also known as the technology time-constant), $V_\text{dd}$ is the power supply voltage amplitude, and $t_\text{tran}$ is the transient time (of a trapezoidal or sinewave clock) which is proportional to the clock period. This Equation \ref{f1} differs from
conventional digital CMOS where the energy per cycle $C V_\text{dd}^2$ is independent of transition time. For AL to consume less power than traditional CMOS, the transition time must exceed $4RC/\alpha$ ($\alpha$
is the switching activity) which poses a challenge for high-performance which requires high clock frequencies. However, as shown in Fig. \ref{adiabatic_trends}, modern FinFET technologies with ps-range RC time-constants can enable significant power savings at GHz frequencies in adiabatic circuits~\cite{liao2014low, yin2025unlocking}.
\section{AC Power Clock Generation}
\begin{figure}[!t]
\centering
\includegraphics[width=3.3in]{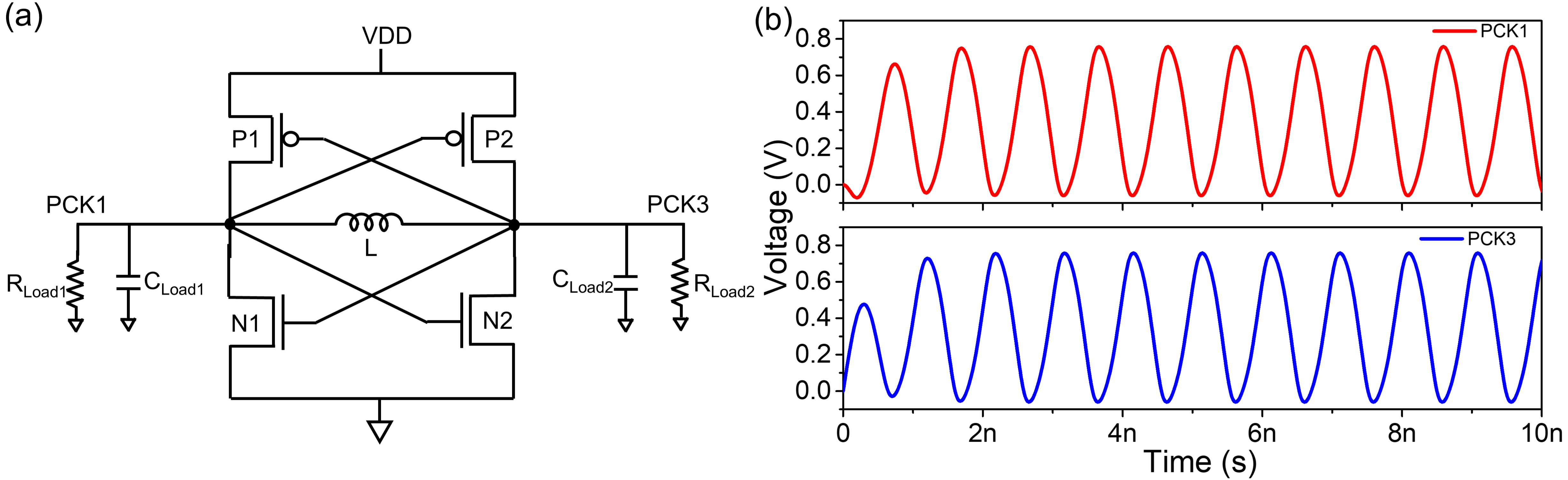}
\caption{(a) The schematic of a resonant 2-phase AC PCK generator with capacitive and resistive loads, and (b) the generated 2-phase sinusoidal PCKs.}
\label{2-phase_gen}
\end{figure}

\begin{figure}[!t]
\centering
\includegraphics[width=3.45in]{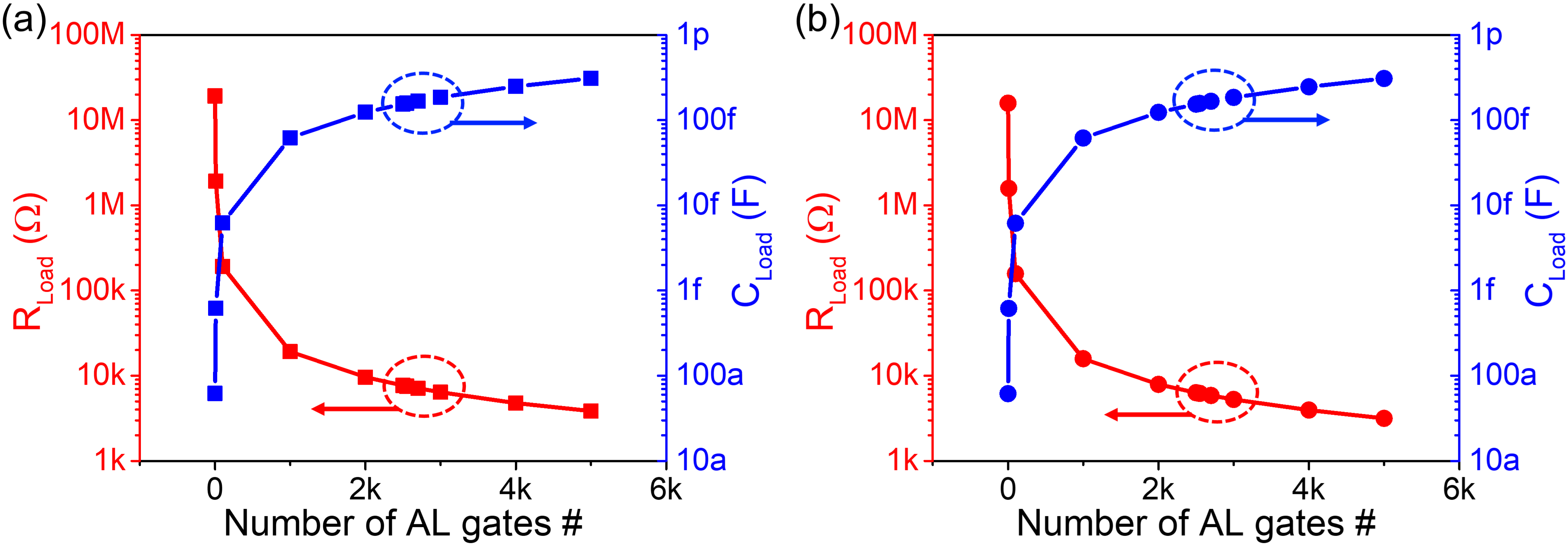}
\caption{R$_\text{Load}$ and C$_\text{Load}$ extraction of (a) PCK1 and (b) PCK3 under ideal AC sources at 1 GHz and 0.8 V.}
\label{idea_AL_gates_RC}
\end{figure}

\begin{figure}[!t]
\centering
\includegraphics[width=3.2in]{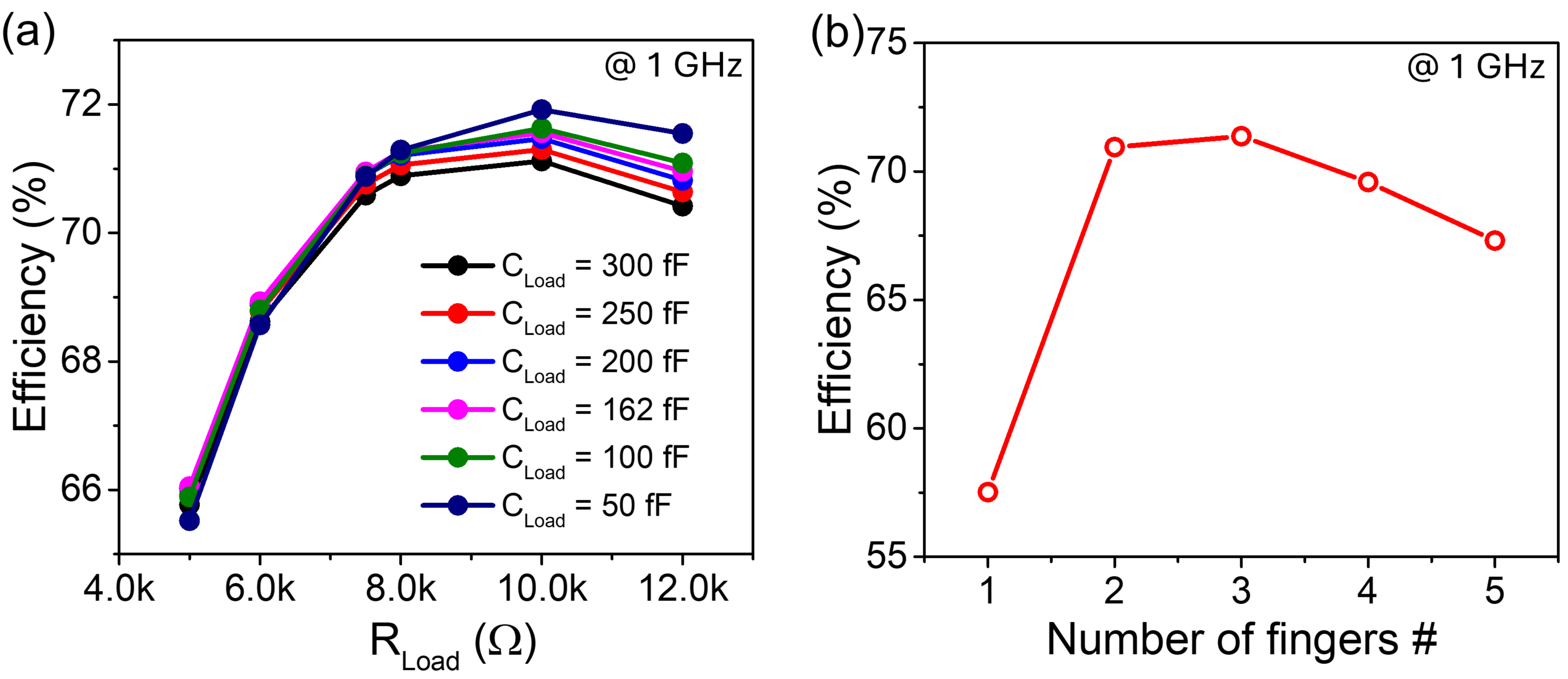}
\caption{(a) The efficiency vs. R$_\text{Load}$ of the 2-phase PCK generator across different C$_\text{Load}$. (b) The efficiency vs. N$_\text{finger}$ with both fixed loads (162 fF, 7.5 k$\Omega$) and 1 GHz.}
\label{2-phase_design_explore2}
\end{figure}

Traditionally AL circuits have used a trapezoidal clock scheme with 4 phases~\cite{lars2018adiabatic} {\em evaluate, hold, recover,} and {\em wait}, but such clocks are hard to generate efficiently, and the  overall power is worse than for other PCK shapes such as blip, or especially sinewave~\cite{ye2001qserl}. 
The number of clock phases is another design consideration, with 1-, 2-, and 4-phase schemes having been explored.~\cite{kim2001true,sasipriya2018design,jeanniot2017synchronised}. 
The AL MAC systolic array uses a 4-phase sinewave AC PCK scheme
with a novel approach to efficiently generating the four synchronized phases. 
The design further eliminates external tank capacitors by using the implicit capacitive loads as part of the clock generator.
This requires a co-design of the circuit and clock generator which complicates the design phase but results in savings in area and power. The building block for the 4-phase PCK is a 2-phase resonant clock generator as shown in Fig.~\ref{2-phase_gen} (a), comprising cross-coupled inverters with a flying inductor L between outputs (PCK1 and PCK3).
With the DC supply VDD, charges oscillate through L between load capacitors C$_\text{Load}$, generating sinusoidal
waves that diminish due to resistive losses. Switches P1 and N2 replenish the dissipated energy
by connecting to VDD and ground at intervals controlled by the resonant circuit itself, producing two 180-degree phase-shifted sinusoidal PCKs at 1 GHz as shown in Fig.~\ref{2-phase_gen} (b). The power delivered to loads is $P_\text{Load}$, while switching losses occur in the four transistors and in the resistive load. Since the oscillation frequency depends on the loads, the resistive and capacitive values of the load must be accurately estimated~\cite{mahmoodi2001efficient,jun24iscas}. Using AC voltage sources driving varying numbers of AL buffers, Fig.~\ref{idea_AL_gates_RC} shows that, as expected, R$_\text{Load1}$ (R$_\text{Load2}$) decreases while C$_\text{Load1}$ (C$_\text{Load2}$) increases with the AL buffer count. For a 2$\times$2 AL MAC systolic array with extracted C$_\text{Load}$ of 162 fF and approximately 2550 AL buffers, the R$_\text{Load}$ is $\sim$7.5 k$\Omega$. Thus, C$_\text{Load1}$ = C$_\text{Load2}$ = 162 fF and R$_\text{Load1}$ = R$_\text{Load2}$ = 7.5 k$\Omega$ are chosen to mimic the AL loads, and the efficiency is calculated as $P_\text{Load}/P_\text{total} \times 100\%$.
Clock generator efficiency depends on C$_\text{Load}$, R$_\text{Load}$, and the N$_\text{finger}$ of FinFETs in the cross-coupled inverter. As shown in Fig.~\ref{2-phase_design_explore2} (a), for fixed C$_\text{Load}$, efficiency initially increases with R$_\text{Load}$ due to decreased conduction losses, then drops as resonant energy transfer becomes under-utilized and reactive current losses dominate. At C$_\text{Load}$ = 162 fF and R$_\text{Load}$ = 7.5 k$\Omega$, efficiency remains above 70\%. Efficiency also increases with N$\text{finger}$ due to increased negative gain for driving loads, as shown in Fig.~\ref{2-phase_design_explore2} (b), but drops with larger N$_\text{finger}$ due to increased switching capacitance and power loss. Since efficiency improves marginally from 2 to 3 fingers, 2 fingers are chosen to minimize area overhead.

\begin{figure}[!t]
\centering
\includegraphics[width=3.3in]{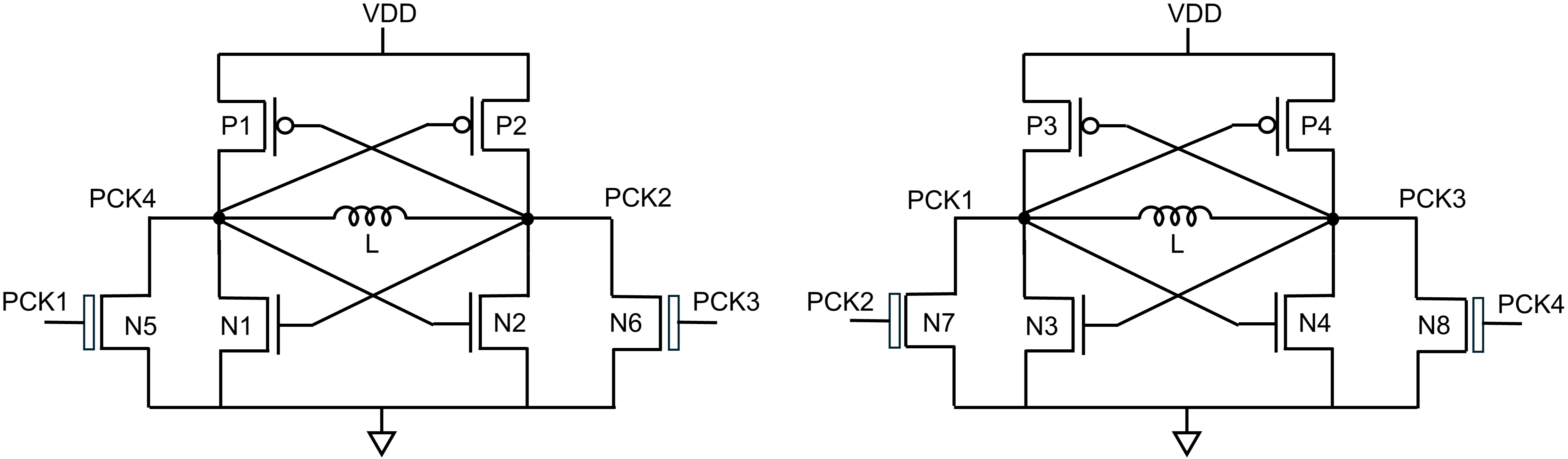}
\caption{The proposed resonant 4-phase AC PCK generator.}
\label{4-phase_PCK_gen}
\end{figure}
\begin{figure}[!t]
\centering
\includegraphics[width=3.1in]{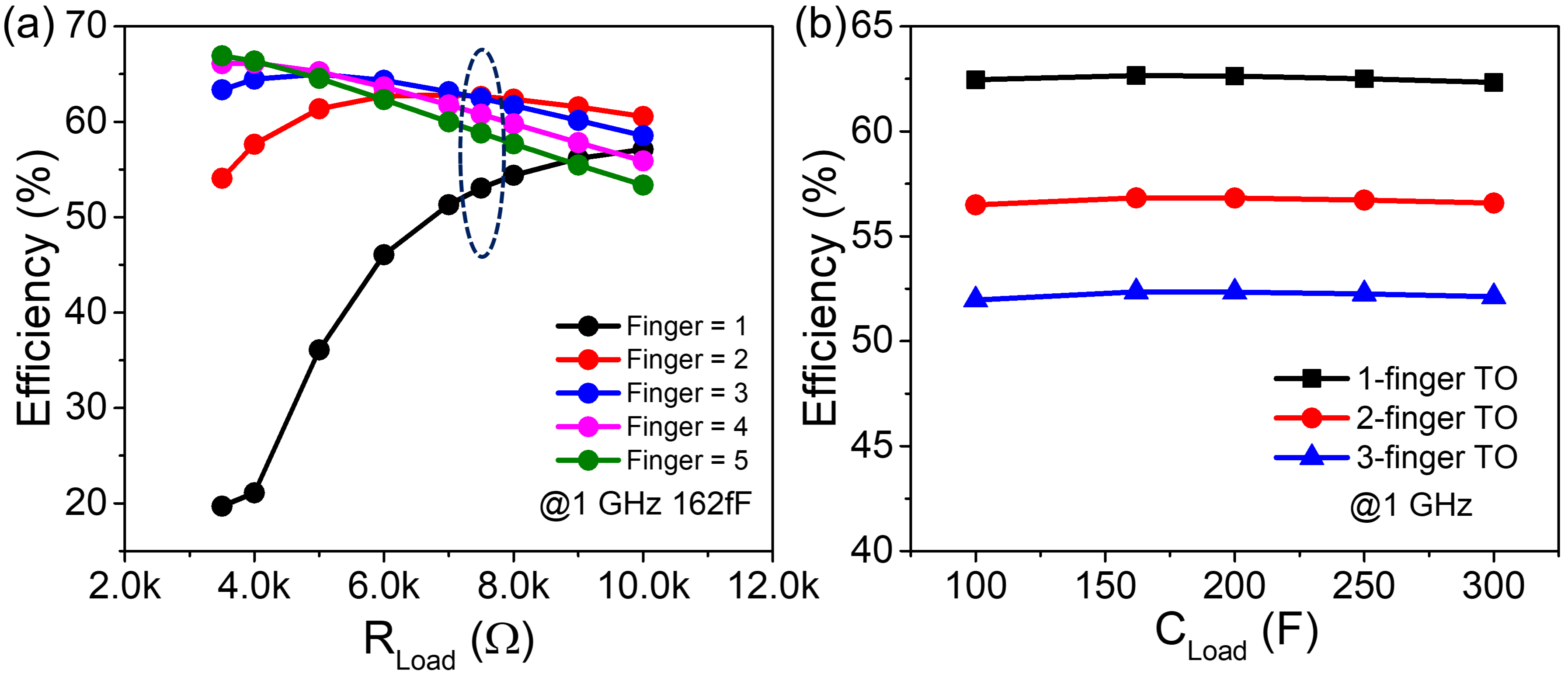}
\caption{(a) The clock efficiency vs. R$_\text{Load}$ under different  N$_\text{finger}$ at a fixed C$_\text{Load}$ of 162 fF and 1 GHz. (b) The clock efficiency vs. C$_\text{Load}$ across varied sizes of TO device at 1 GHz.}
\label{4-phase_design_explore2}
\end{figure}

To generate 4-phase PCK with the 2-phase clock generator, as shown in Fig.~\ref{4-phase_PCK_gen}, two sets of 2-phase clock generators are required. In addition, they also need shunt NFETs at each output node, with the gate terminals coupling from the next phase outputs ~\cite{ahn20035}, to shift the phase of PCK by 90 degrees.
\begin{figure}[!t]
\centering
\includegraphics[width=3.25in]{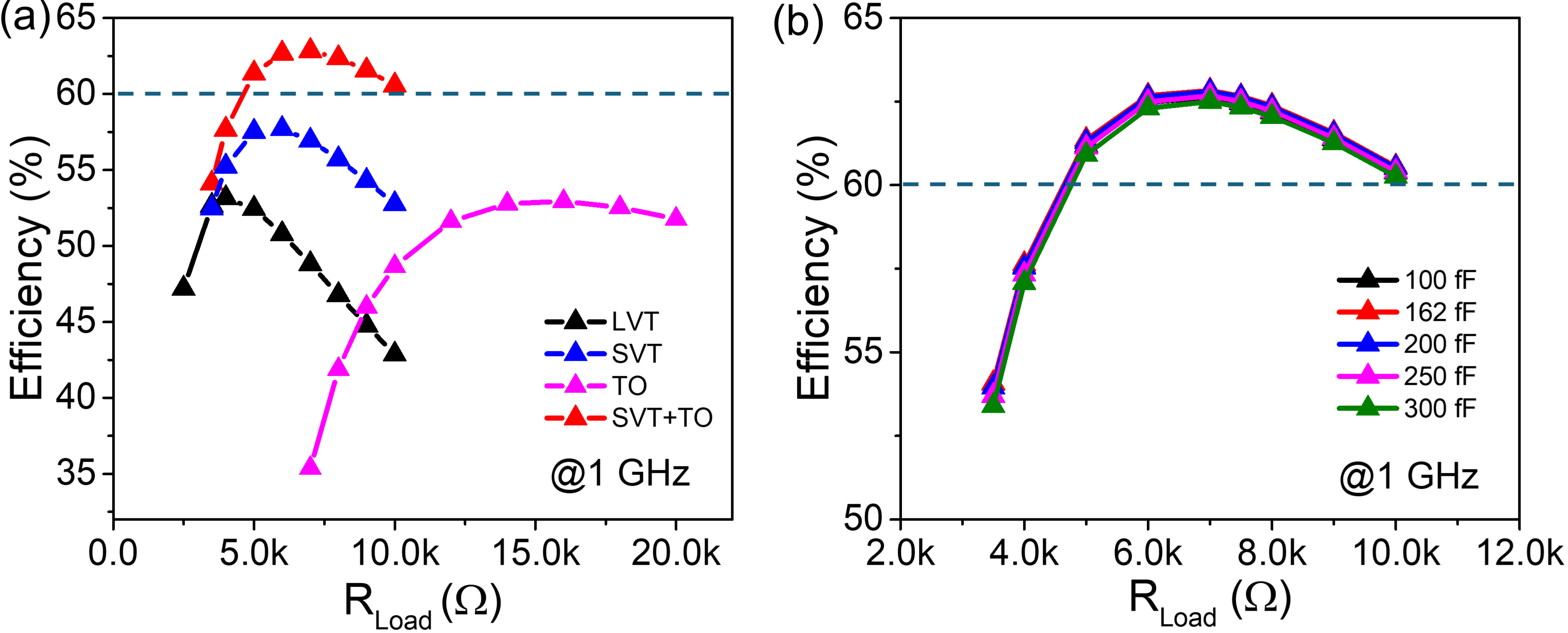}
\caption{(a) The clock efficiency vs. R$_\text{Load}$ of the 4-phase PCK generator based on LVT, SVT, TO, and SVT+TO device under a fixed C$_\text{Load}$ of 162 fF and frequency of 1 GHz. (b) The clock efficiency vs. R$_\text{Load}$ across different C$_\text{Load}$ at 1 GHz.}
\label{4-phase_design_explore}
\end{figure}
Fig.~\ref{4-phase_design_explore} (a) shows 4-phase clock efficiency under different R$_\text{Load}$ with fixed C$_\text{Load}$ = 162 fF and 1 GHz frequency. Using SVT devices in cross-coupled inverters and thick oxide (TO) devices in shunt NFETs, i.e., SVT+TO, the optimized peak efficiency is 62.5\% at 7 k$\Omega$, remaining above 60\% at 7.5 k$\Omega$. Non-mixed options (LVT only, SVT only, TO only) achieve peak efficiency much less than 60\%, making SVT+TO the optimal choice. SVT transistors in the cross-coupled structure provide sufficient negative gain, while TO NFETs as shunt devices reduce power consumption. Across various C$_\text{Load}$ from 100 fF to 300 fF shown in Fig.~\ref{4-phase_design_explore} (b), the SVT+TO clock generator maintains efficiency above 60\% with marginal changes at 1 GHz, delivering sufficient power to AL circuits under limited input power budget. Fig.~\ref{4-phase_design_explore2} (a) shows efficiency versus R$_\text{Load}$ across different FinFET N$_\text{finger}$ (N1-N4 and P1-P4) in the cross-coupled inverter, with NFET to PFET size ratio 1:1. At targeted AL load 162 fF, efficiency is sensitive to small R$_\text{Load}$ and small FinFET size because insufficient gain causes active losses. Efficiency becomes less sensitive as size increases. Optimal SVT FinFET size is 2 fingers at R$_\text{Load}$ = 7.5 k$\Omega$ (marked by dark blue circle). With SVT FinFET size fixed, shunt TO size must be optimized to minimize power loss. Fig.~\ref{4-phase_design_explore2} (b) shows that larger TO devices reduce efficiency with the shunt NFET dominating internal losses. TO size should remain minimal or increase to 2 fingers ($>56\%$) only if larger negative gain is needed for heavy loads.
\section{DESIGN of AL MAC SYSTOLIC ARRAY}
\subsection{AL Complex Digital Computing Logic}

\begin{figure}[!t]
\centering
\includegraphics[width=2.5in]{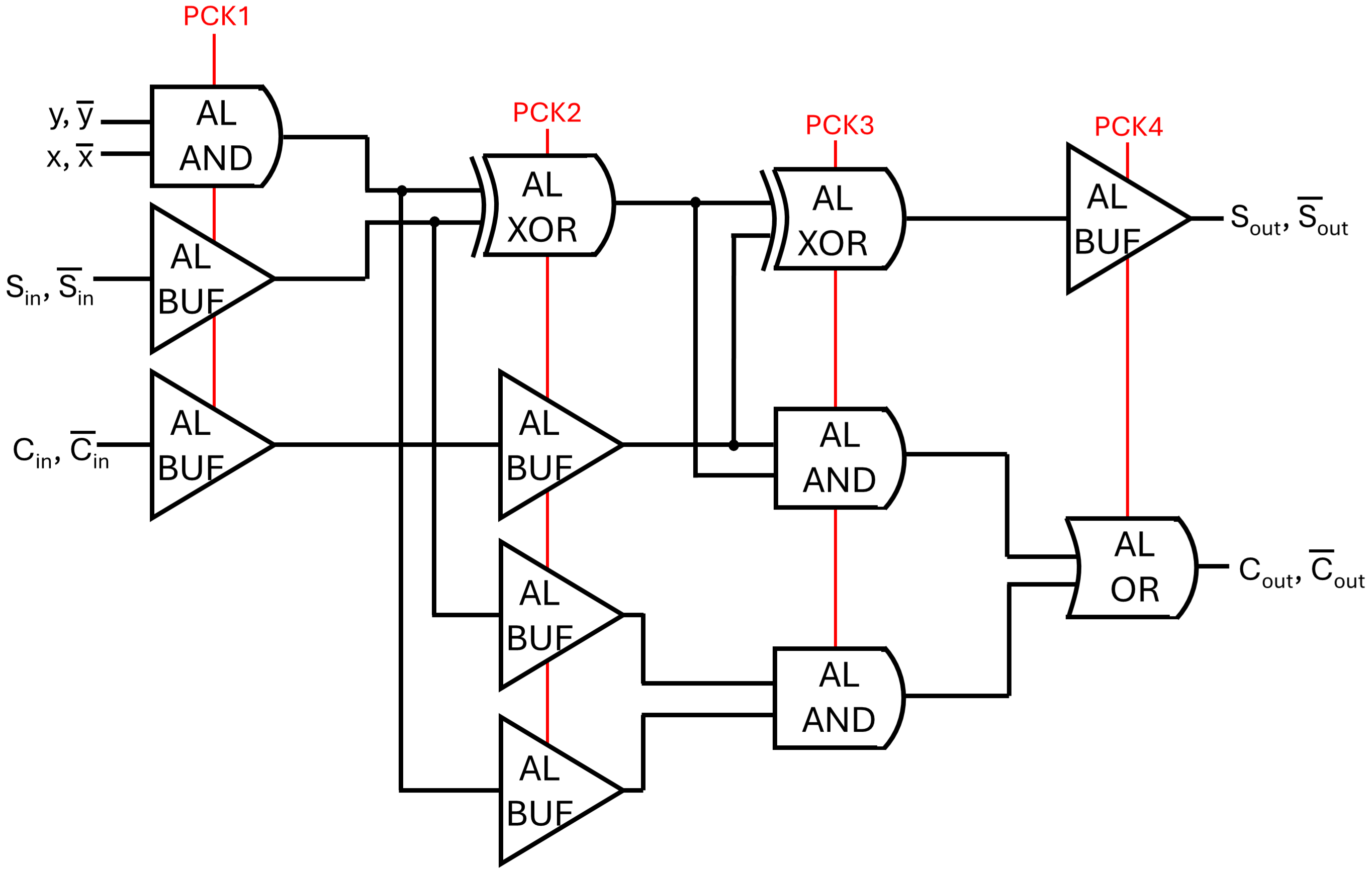}
\caption{The gate-level schematic of AL CSA with PCKs.}
\label{AL_CSA}
\end{figure}

\begin{figure}[!t]
\centering
\includegraphics[width=3.2in]{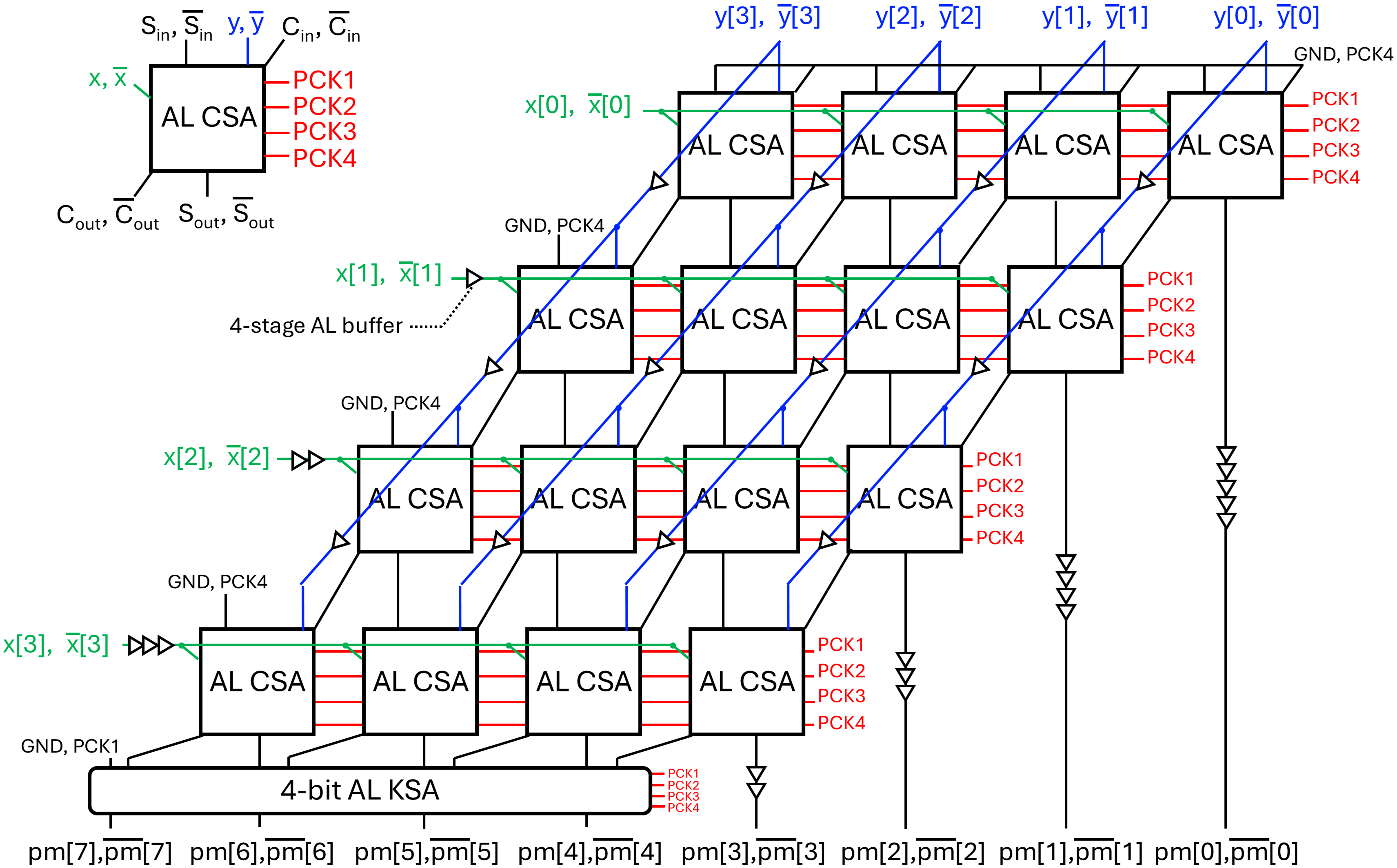}
\caption{The gate-level schematic of AL multiplier.}
\label{AL_multipler}
\end{figure}

The core of our AL multiplier with shift-and-add functionality is the AL CSA shown in Fig.~\ref{AL_CSA}, where \textit{dual-rail complementary signals} (e.g., $x$ and $\bar{x}$) \textit{are transmitted on single wires} 
(this convention applies to all subsequent figures). 
Unlike traditional digital CSA combining an AND gate with a full adder (FA), all AL FA inputs and their complements must be buffered with AL buffers (along with all gates inside the AL FA) in order to synchronize the AL signals through the 4-phase PCK scheme. This buffering ensures: 1) the number of power supplies for the AL CSA is an exact multiple of 4, enabling array duplication without phase ordering concerns; 2) each AL CSA output synchronizes with the last PCK phase. A 4-bit operand word length is targeted to balance design complexity for this proof of concept.
As shown in Fig.~\ref{AL_multipler}, a 4$\times$4 AL CSA array is constructed to perform shift and add multiplications. Since the data flow is two-dimensional, i.e., X-axis direction (row-wise) and Y-axis direction (column-wise), and the delay between inputs and output after a set of 4-phase PCK
sequence is one clock cycle, the input and output data bits should be handled carefully such that the row-wise input data bits arrive at the same time as the intermediate column-wise outputs from the previous AL CSA. Additionally, all AL outputs of the AL multiplier should be output simultaneously. To achieve the input and output synchronization, serial AL buffers are inserted at the row-wise inputs and column-wise outputs by counting how many sets of 4-phase PCKs the AL signals pass through. For example, from the left side of the AL CSA array, the row inputs of row 0 should not be buffered, while the inputs of other rows need to be buffered by increasing the number of AL buffers with a step of 4, i.e., 4 AL buffers at the left inputs of row 1, 8 AL buffers at the left inputs of row 2, and 12 AL buffers at the left inputs of row 3. Similarly, output columns 0--3 require AL buffers while columns 4--6 do not, ensuring that row-wise inputs align with intermediate carry and sum signals, and all outputs are synchronized under the 4-phase PCK scheme.

\begin{figure}[!t]
\centering
\includegraphics[width=3.35in]{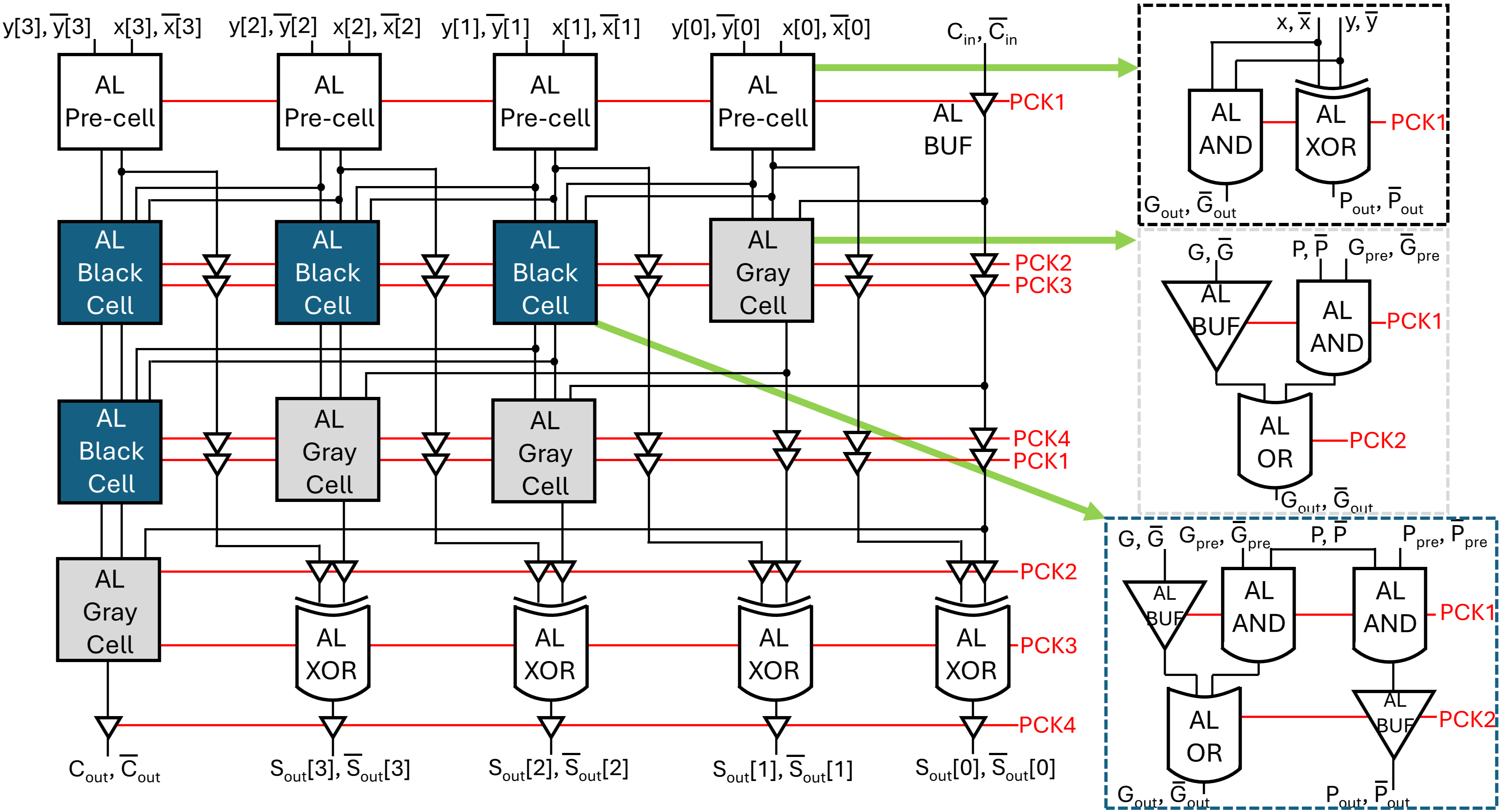}
\caption{Schematic of a 4-bit AL KSA, including AL black-cells, AL pre-cells, AL gray-cells, AL XOR, and AL buffers.}
\label{AL_KSA_4bits}
\end{figure}

\begin{figure}[!t]
\centering
\includegraphics[width=3.5in]{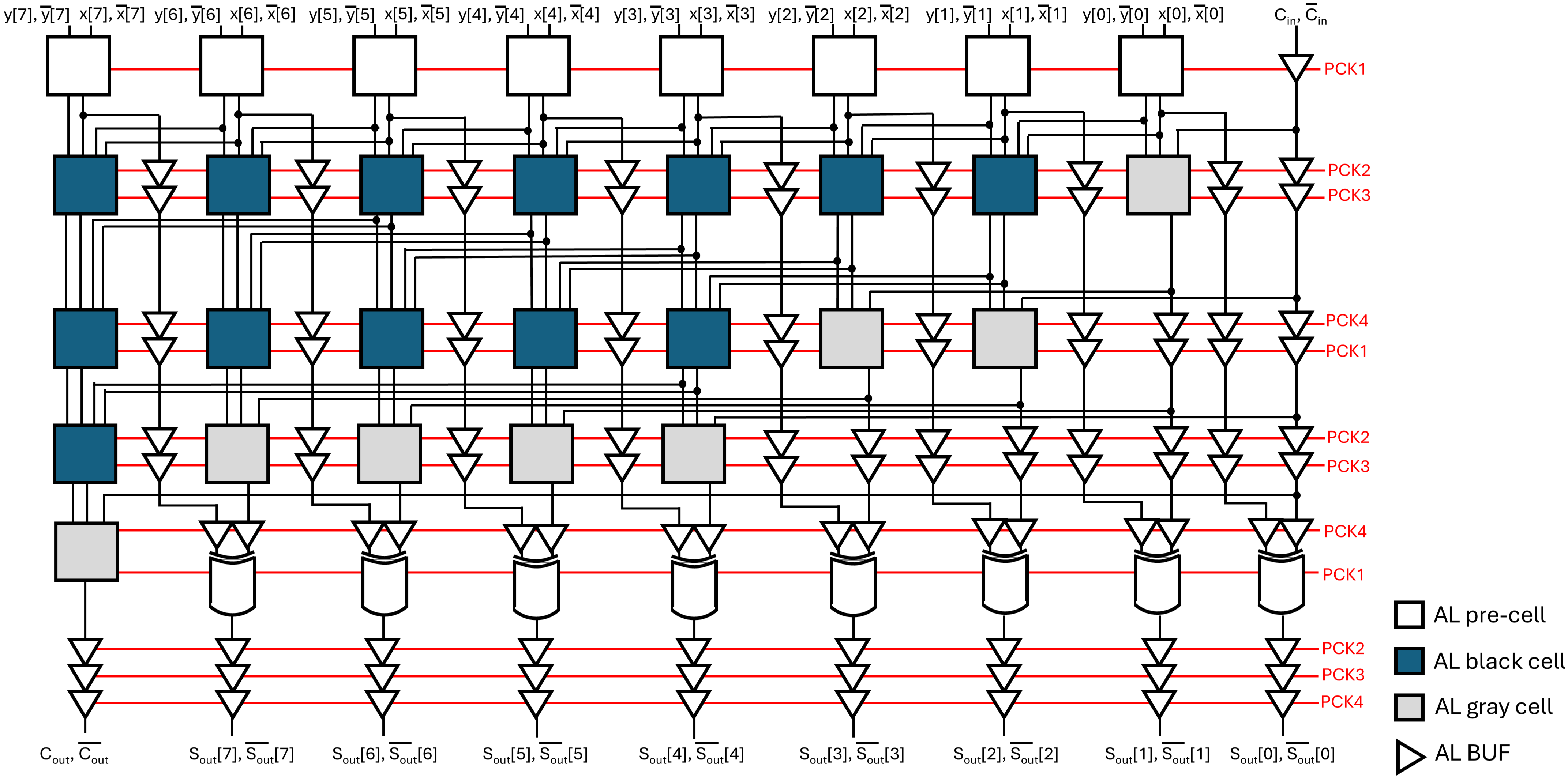}
\caption{The schematic of a 8-bit AL KSA, including AL black-cells, AL pre-cells, AL gray-cells, AL XOR, and AL buffers.}
\label{AL_KSA_8bits}
\end{figure}
A 4-bit AL KSA adds the carry-outs from the most significant bits (MSBs) of AL CSA array to their corresponding sum outputs. The 4-bit AL KSA comprises AL pre-stage cells (combining AL XOR and AND gates to generate $P$ and $G$ signals), AL black cells (combining $P$ and $G$ from adjacent positions using two AL AND gates and an AL OR gate), and AL grey cells (computing only $G$ using an AL AND gate and an AL OR gate), as shown in Fig.~\ref{AL_KSA_4bits}. The AL black cells enable parallel prefix logic by viewing bits from the right and extending carry look-ahead distance. For example, bit 2 combines with bit 0, and bit 3 with bit 1, ensuring carry-in information from the LSB is processed in parallel without ripple delay. AL grey cells in the final stage complete the carry computation for all bits. The final step computes sum bits using AL XOR gates that take the $P$ signal and corresponding carry-in as inputs, performed simultaneously for all 4 bits. AL buffers synchronize internal signals throughout each cell. Thus, a 4$\times$4 AL CSA array and 4-bit AL KSA with buffered inputs and outputs comprise
a 4-bit AL multiplier.

For the AL accumulation circuits, an 8-bit AL KSA, as shown in Fig.~\ref{AL_KSA_8bits}, is designed to implement the addition function. The design methodology of the 8-bit AL KSA is similar to that of the 4-bit AL KSA mentioned before and will not be described further.
\begin{figure}[!t]
\centering
\includegraphics[width=3.2in]{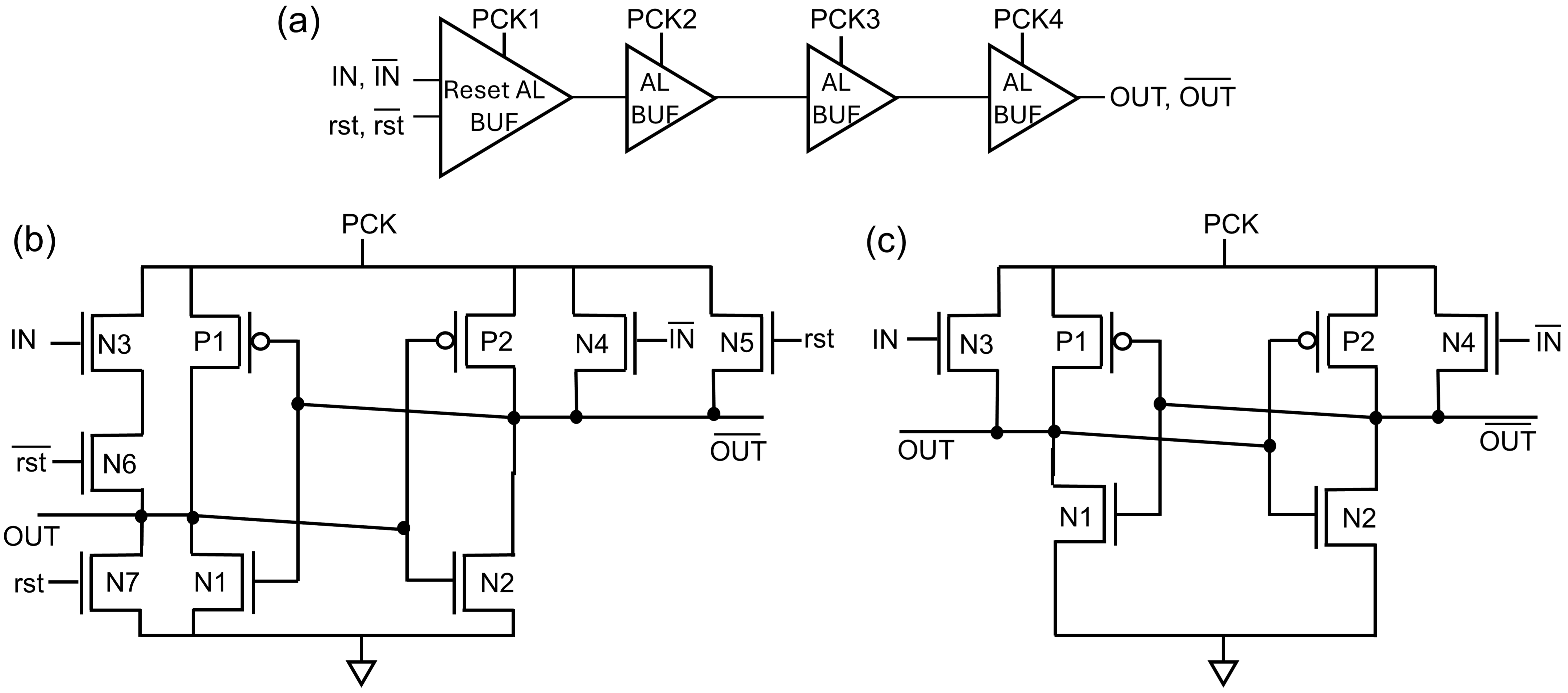}
\caption{The gate-level schematic of (a) AL delay element, and transistor-level schematic of (b) AL buffer with reset function and (c) AL buffer.}
\label{AL_delay_ele}
\end{figure}
To achieve a
pipeline functionality the key element is the AL delay element. Unlike traditional flip-flops, which use the clock to hold the data from the previous clock cycle, the AL delay element takes advantage of the 1 clock cycle delay between AL inputs and outputs when implementing the 4-phase PCK scheme. As shown in Fig.~\ref{AL_delay_ele} (a), the structure of the AL delay element is simple: 4 stages of AL buffers in serial connection, with the first AL buffer, as shown in Fig.~\ref{AL_delay_ele} (b), inserting reset inputs to reset the initial value to 0 before starting the calculation. The
next 3 stages are regular AL buffers with a structure shown in Fig.~\ref{AL_delay_ele} (c). Therefore, the AL accumulation
can be achieved as follows: the product of the AL multiplier is fed to one input bus of the 8-bit AL KSA, while another input bus is reset to all 0s by the reset function of the AL delay element when the accumulation circuits start to accumulate for the first time. After the first
summation, the accumulated results are delayed by one clock cycle, such that they can arrive at the input bus at the same time as the next multiplied product.
\subsection{AL MAC PE Design}

\begin{figure}[!t]
\centering
\includegraphics[width=3.3in]{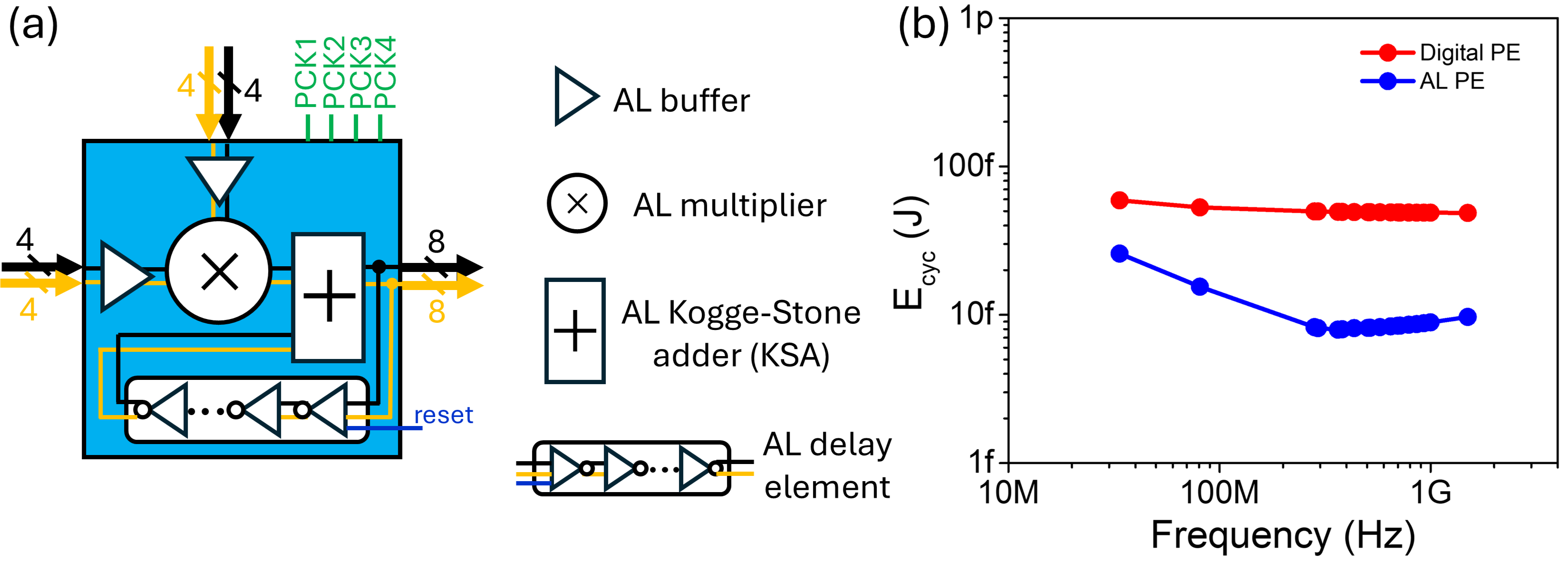}
\caption{(a) AL MAC PE schematic and (b) $E_\text{cyc}$ comparison between a digital PE and an AL PE across varied frequencies.}
\label{AL_PE_1GHz}
\end{figure}

As shown in Fig.~\ref{AL_PE_1GHz} (a), a 4-bit AL MAC PE includes AL buffers, a 4-bit AL multiplier, an 8-bit AL KSA, and AL delay elements with reset features. Each mentioned component is constructed by PFAL gates, and then its $E_\text{cyc}$ is compared to that of the digital MAC PE constructed with similar digital logic gates. Fig.~\ref{AL_PE_1GHz} (b) shows a similar tendency to \cite{amirante2001variations} that $E_\text{cyc}$ of proposed AL PE, with a minimum value across frequencies, is lower than that of the digital PE at 0.8 V. The $E_\text{cyc}$ of AL PE is reduced by 50\% compared to that of the digital PE at 1 GHz, which is close to the minimum $E_\text{cyc}$. It is notable that this approach not only lowers power consumption but also enables higher frequencies without compromising performance.

\subsection{Signal Conversion Circuits}
\begin{figure}[!t]
\centering
\includegraphics[width=3.3in]{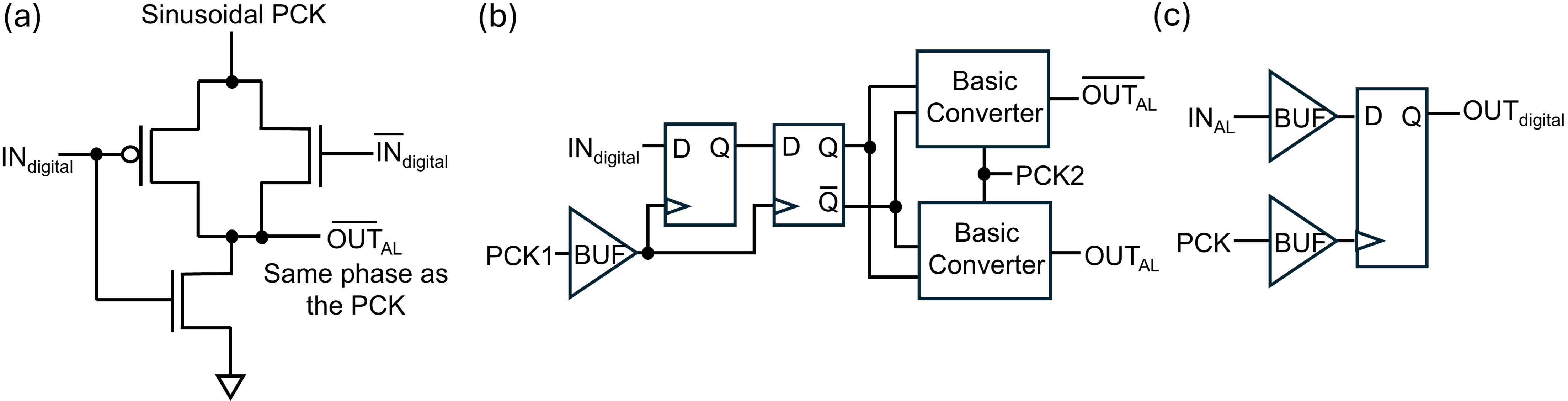}
\caption{(a) Basic converter; (b) Digital-to-AL conversion (DALC) circuits; (c) AL-to-digital conversion (ALDC) circuits.}
\label{DALC_converter}
\end{figure}

Since the AL MAC PE core receives and outputs AL signals, it needs digital-to-AL conversion (DALC) circuits at its inputs and AL-to-digital circuits at its outputs to interact with memories, i.e., first-in-first-out (FIFOs) circuits and registers, which require digital inputs and outputs for applications. To achieve DALC, a basic converter~\cite{fischer2003adiabatic} converts rectangular digital signals to sinusoidal using one of the 4-phase PCK supplies as shown in Fig.~\ref{DALC_converter} (a). For AL inputs, transmission gates serve as the most effective pull-up structure since sinusoidal PCK passes through without distortion or threshold-induced voltage steps; gate sizing follows digital circuit principles, with energy dissipation dominated by resistance rather than delay. Since AL MAC PE core inputs are dual rails, the entire DALC schematics shown in Fig.~\ref{DALC_converter} (b) requires two basic converters to encode digital signals with PCK phase 1 to complementary input signals. Inputs are gated by serial-connected digital registers to avoid metastability, and a digital inverter converts PCK phase 0 to a clock for positive-edge-triggered registers. The digital input must remain stable during precharge, hold, and recovery phases, switching only during the wait phase when PCK is at ground. AL-to-digital conversion (ALDC) circuits at the AL MAC PE core outputs, shown in Fig.~\ref{DALC_converter} (c), translate AL to digital signals by adding registers to the core output, using the hold phase of PCK as the digital clock.

\subsection{Architecture of AL MAC Systolic Array}\label{matrix_cal}

\begin{figure}[!t]
\centering
\includegraphics[width=2.8in]{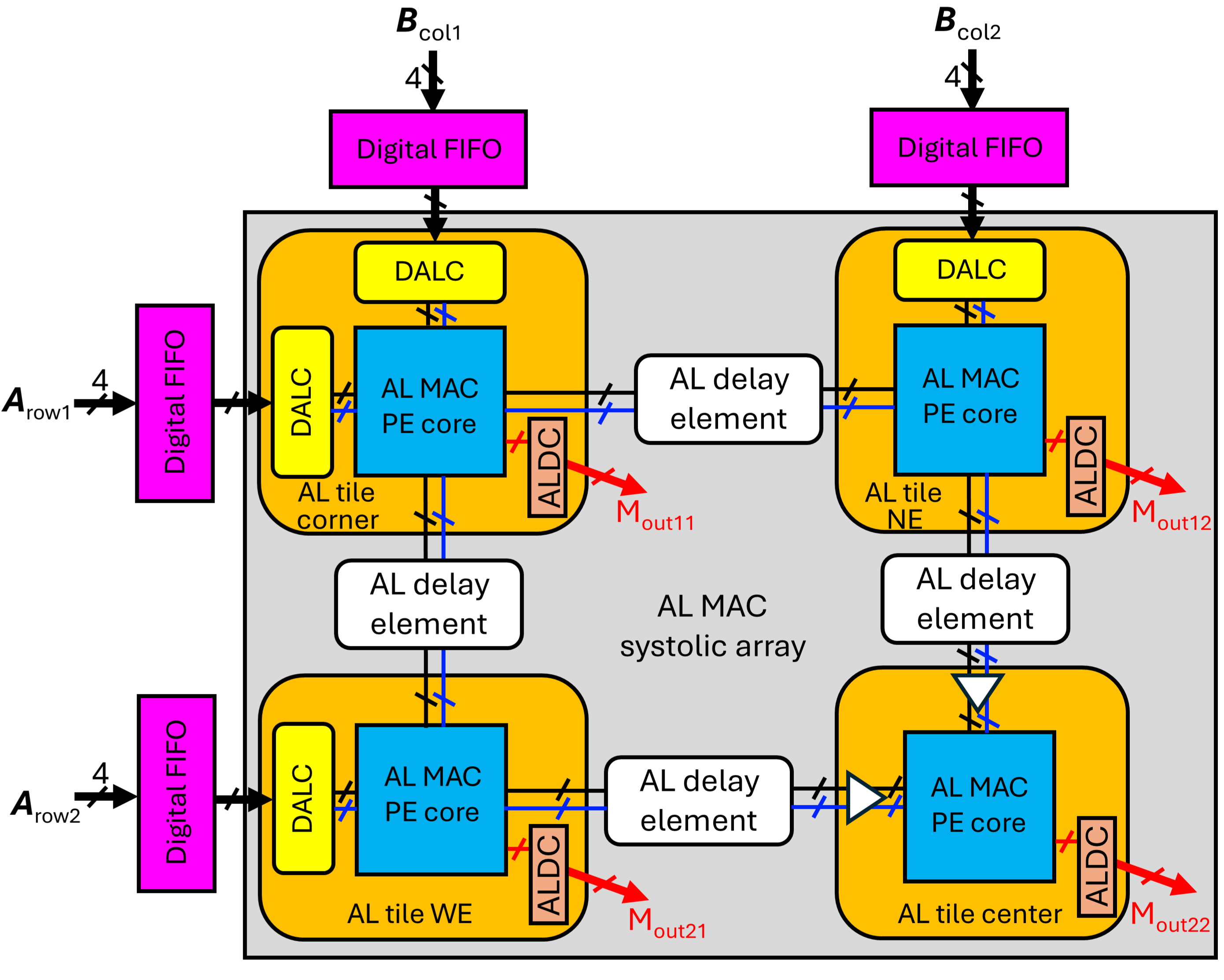}
\caption{Hierarchical architecture of $2\times2$ AL MAC systolic array. Black and blue lines represent dual rail AL buses.}
\label{2x2_AL_systolic_array}
\end{figure}

Fig.~\ref{2x2_AL_systolic_array} shows the AL MAC systolic array architecture for matrix multiplications, requiring four 4-bit input digital FIFOs: 2 for row-wise and 2 for column-wise data streams. Unlike normal digital FIFOs that output data at each clock cycle, when interacting with AL circuits, digital inputs must remain stable throughout all 4 phases. The modified FIFO takes data serially each clock cycle but holds data for 4 cycles when driving the DALC, with a resettable counter tracking hold cycles and resetting at 4. Additionally, input data must be preprocessed: for matrix multiplication $\mathbf{A}_{2\times n}\times\mathbf{B}_{n\times 2}$ producing $\mathbf{C}_{2\times 2}$, zeros are added at the beginning of the second row of $\mathbf{A}$ and second column of $\mathbf{B}$ to ensure correct pipelining. The FIFOs are synthesized using Synopsys DC compiler with modified 4-bit FIFO RTL mapped to 16 nm FinFET technology.
An AL MAC PE core with peripheral signal conversion circuits forms an AL MAC tile. A 2$\times$2 AL MAC systolic array contains 4 AL tiles with AL delay elements between adjacent tiles for pipelining. Tiles are categorized into four types based on DALC count, optimized for area and energy efficiency. The AL tile at the top-left corner (AL tile corner) requires both row-wise and column-wise DALCs to convert digital inputs to AL inputs for the AL MAC PE core. After 4 clock cycles of pipelining by AL delay elements, the AL tiles at the northern edge (AL tile NE) and western edge (AL tile WE) begin calculation; the AL tile NE requires only column-wise DALCs since row-wise data is already converted to AL signals, while the AL tile WE requires only row-wise DALCs. The AL tile at the bottom-right corner or center (AL tile center) receives AL data from both directions and contains no DALCs, only AL buffers to synchronize signals with the PCK phase. This architecture allows the AL tile center, AL tile NE, and AL tile WE to be duplicated multiple times for larger systolic arrays.
\section{Simulation Results}
\begin{figure}[!t]
\centering
\includegraphics[width=3.5in]{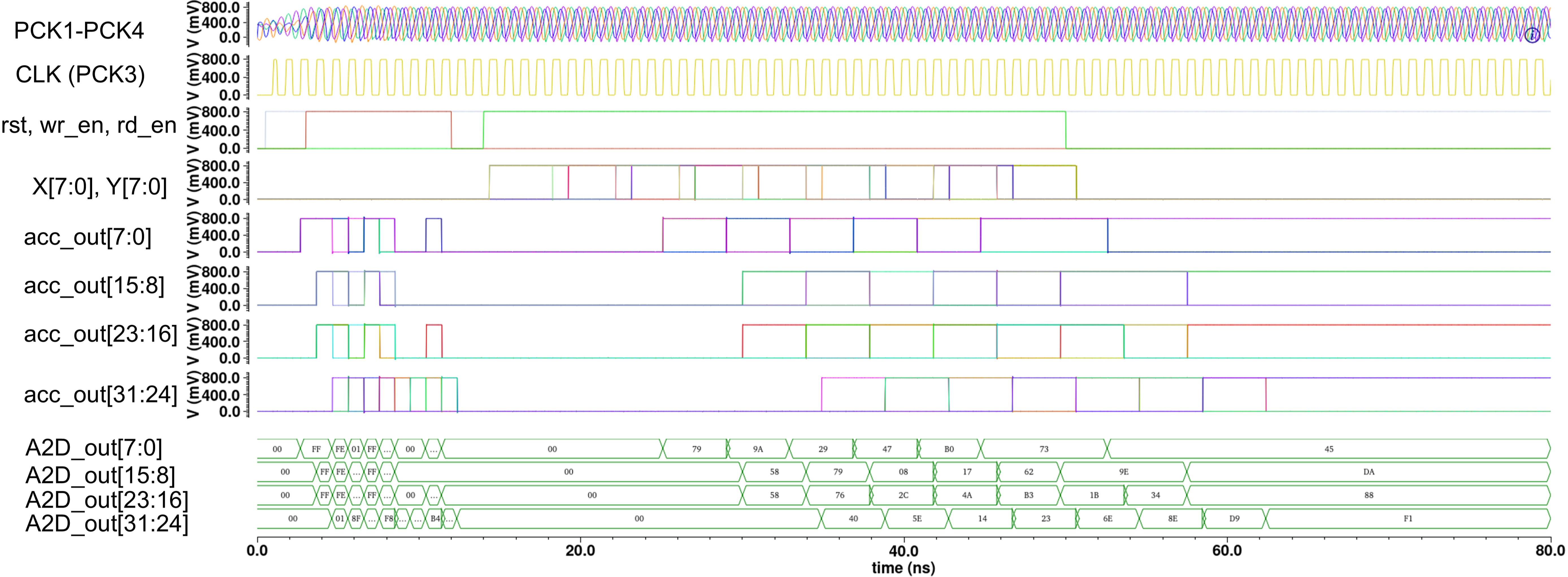}
\caption{The input and output waveforms of the AL MAC systolic array with digital interfaces.}
\label{AL_wave_form}
\end{figure}
The designed 4-phase clock generator was integrated with the entire 2$\times$2 AL MAC systolic array to provide valid 4-phase PCKs. Only one DC 0.8V voltage source is used as power supply. The simulation results for functional checking are shown in Fig. \ref{AL_wave_form}. Before the data stream inputs, the reset (rst) is set to 0 to initialize the value of the entire array to all 0s.

\begin{table*}[!t]
  \centering
\caption{Comparison with SotA accelerators}
\label{AL_comparison2}

  \renewcommand{\arraystretch}{0.9}
  \resizebox{\textwidth}{!}{
  \begin{tabular}{lccccc}
    \toprule
    \textbf{Metrics} & \textbf{2021 ISCA \cite{jouppi2021ten}} & \textbf{2023 TCAD \cite{yang2022dtatrans}} & \textbf{2025 TC \cite{wang2025desa}} & \textbf{2023 JSSC \cite{DIANA-JSSC-2023}} & \textbf{This work} \\
    \midrule
    Architecture & Systolic & Systolic &  Systolic & Spatial & Systolic \\
    Function & Digital MAC &  Digital MAC & Digital MAC & Digital MAC & AL MAC \\
    Design Level & System-level & Core-level  & Core-level & Core-level & System-level \\
    Technology & 7 nm & 40 nm & 16 nm & 22 nm & 16 nm \\
    Precision & INT8 & INT4 & INT8 & INT8 & INT4 \\
    Array Size & 4$\times$128$\times$128 & 476 & 4288 & 256 & 16$\times$16 \\
    Frequency (GHz) & 1.05 & 1  & 0.2 & 0.28 & 1 \\
    Power (mW) & 175000 & 803 &   10990 & 132.2 & 16.04 \\
    Energy Efficiency (TOPS/W) & 0.786 & 1.623  & 0.072 &1.74 & 7.98 \\
    Norm. Power (mW) & 261 & 189.3  & 164 & 22.04 & \textbf{16.04} \\
    Norm. Energy Efficiency (TOPS/W) & 1.964 & 2.705  & 0.624 & 6.49 & \textbf{7.98} \\
    \bottomrule
  \end{tabular}
  }
\end{table*}

\subsection{End-to-End Functional Validation}
As the clock generator begins
oscillation, rst is set to 1. Synthesized digital FIFOs are used to write data from the input data stream when the write enable (wr\_en) is on, while reading data to output data stream every 4 clock cycles to the DALC modules when the read enable (rd\_en) is on. The structure of data inputs is two matrices: $\mathbf{X}$ is 2 rows by 8 columns and $\mathbf{Y}$ is 8 rows by 2 columns corresponding to the matrix $\mathbf{A}$ and $\mathbf{B}$  mentioned in Section \ref{matrix_cal}, respectively. The elements in each matrix are 4-bit integers. A digital clock converted by ALDC tracks the phase of PCK3 to synchronize the digital inputs and outputs, and has a frequency of 1 GHz. After the data pass through the AL MAC systolic array, the calculated results are pipelined out correctly
and the values are held unchanged when rd\_en is off.

\begin{table}[!t]
  \centering
\caption{Power Comparison between AL and Digital System}
\label{AL_comparison}
\begin{small}
\begin{tabular}{cc}
    \toprule
    Section & Power $\mu$W @ 1 GHz\\
    \midrule
    Pure AL PEs total & 38.1 \\
    AL systolic MAC core & 174.4 \\
    Power supply circuits & 102.7 \\
    Total delivered power & \textbf{537.1}\\
    \hline
    Digital systolic MAC core & 302.2\\
    Digital total delivered power & 839.5 \\
    \hline 
    Power saving on systolic MAC-core level & \textbf{42\%} \\
    Power saving on the system level & \textbf{36\%} \\
  \bottomrule
\end{tabular}
\end{small}
\end{table}

\subsection{System-Level Evaluation of the AL Systolic Array}

Table~\ref{AL_comparison} shows a power consumption comparison between the 2$\times$2 AL systolic array and its digital counterpart at the MAC core level, and at the system-level (including digital FIFOs). At the MAC core level, the AL systolic MAC core shows a power consumption of 174.4 $\mu$W for a reduction of 42\% compared to that of the digital systolic MAC core. Notably, the total power consumption of the pure AL PEs (excluding converters) is only 38.1 $\mu$W. Thanks to our efficient 4-phase AC clock generator, the generator consumes 102.7 $\mu$W, and thus the total delivered power, i.e., the AL system including AL MAC systolic array and digital peripherals, is 537.1 $\mu$W. Compared to the digital power of 839.5 $\mu$W, the AL system gives a very promising power saving of 36\% at 1 GHz.

\begin{figure}[!t]
\centering
\includegraphics[width=2.2in]{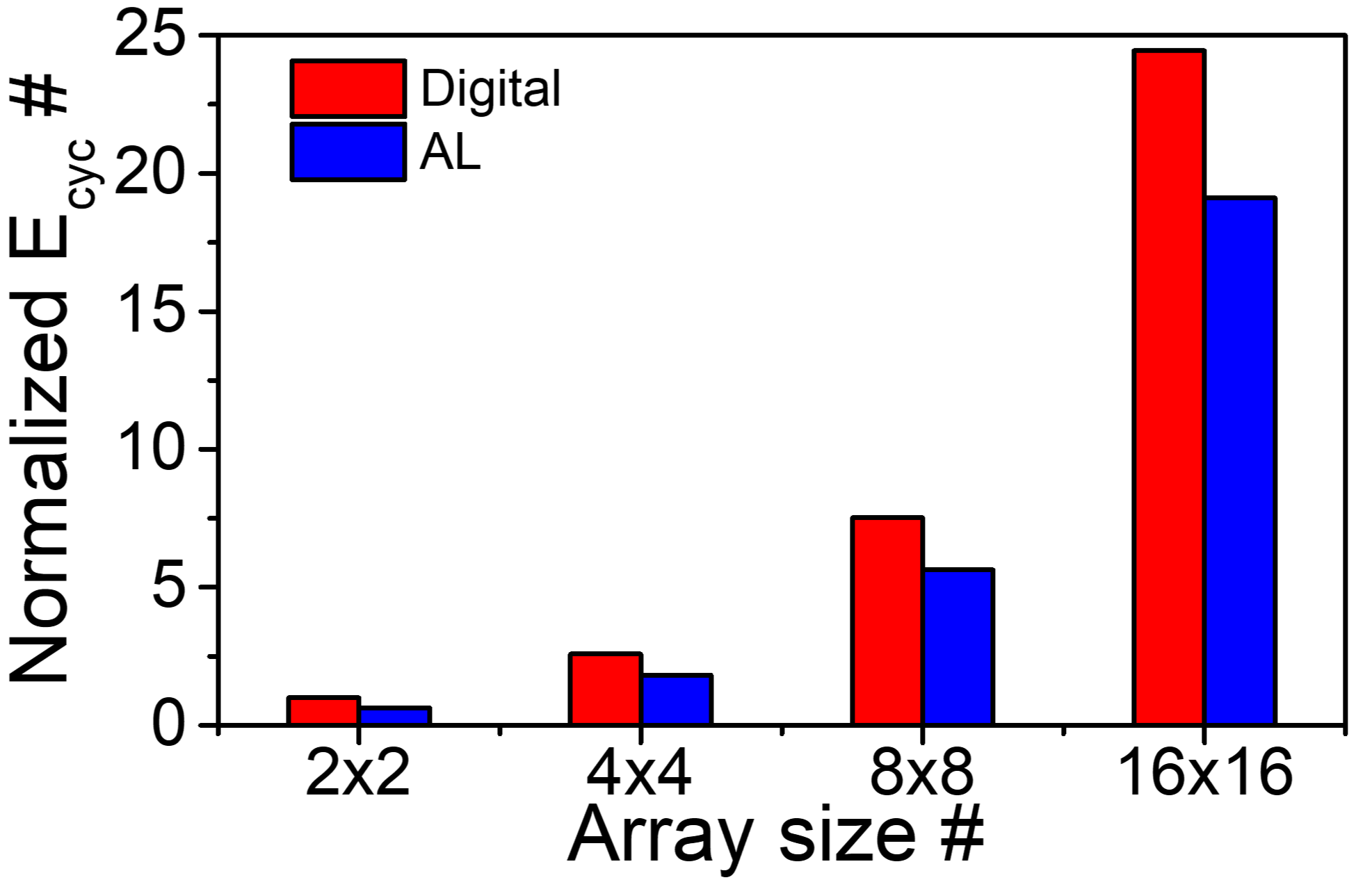}
\caption{Normalized E$_\text{cyc}$ comparison of the digital and AL MAC systolic array system for various array sizes at 1 GHz.}
\label{estimation_scaling}
\end{figure}

For scaling up the size of the AL MAC systolic array, the system architecture must recognize the strengths and constraints of AL. Given the power consumption of the 2$\times$2 digital and AL MAC systolic arrays at 1 GHz and 0.8 V, the E$_\text{cyc}$ of various size systolic arrays can be estimated to provide further insights.
In Fig.~\ref{estimation_scaling} the E$_\text{cyc}$ of the 2$\times$2 digital MAC systolic array with peripherals is set as baseline while the E$_\text{cyc}$ of other designs with different sizes is normalized based on it. As the array size increases, the E$_\text{cyc}$ of both digital and AL MAC systolic arrays increases. Notably, the AL MAC systolic array with peripherals always achieves lower E$_\text{cyc}$, and it can help reduce the  $E_\text{cyc}$ by $\sim$22\% at a large size of 16$\times$16.

Table \ref{AL_comparison2} shows a comparison of projected normalized metrics between the proposed AL systolic array and SotA digital accelerators from the literature. The power and energy efficiency are normalized to 16 nm and a 16$\times$16 array size with INT4 throughput. Technology normalization follows DeepScaleTool \cite{sarangi2021deepscaletool}, which provides technology-scaling factors for estimating equivalent power across CMOS nodes. For precision normalization, we assume that MAC datapath cost scales approximately quadratically with operand bit width; therefore, one INT8 MAC is treated as equivalent to four INT4 MACs. With a normalized power consumption of 16.04 mW, our proposed AL solution achieves the lowest power footprint across all other designs, representing a 1.37$\times$ reduction compared to the next most efficient design \cite{DIANA-JSSC-2023} and a 16.3$\times$ improvement over \cite{jouppi2021ten}. The projected normalized energy efficiency of 7.98 TOPS/W surpasses all SotA architectures and provides a strong argument for AL to be implemented in Silicon in advanced FinFET nodes.




\subsection{Clock-Generator Non-Ideality Analysis}



We evaluate finite inductor quality factor by adding series resistance to the resonant inductor at 1~GHz. With 4-finger devices in the cross-coupled inverter pairs, the clock generator maintains stable four-phase power-clock generation and correct AL systolic-array operation down to an effective quality factor of approximately 14. Increasing the cross-coupled devices to 6 fingers provides stronger negative resistance, allowing the generator to remain functional at an effective quality factor of approximately 9.8. These results indicate that finite-$Q$ degradation primarily reduces the oscillation margin, and that the additional tank loss can be partially compensated by increasing the negative resistance of the clock generator.


\subsection{Voltage/Frequency Operating Region}

To characterize the AL datapath margin, we sweep the ideal four-phase PCK amplitude and operating frequency. This isolates the intrinsic voltage/frequency limit of the AL systolic array from the startup limit of the resonant clock generator.

\begin{figure}[t]
    \centering
    \includegraphics[width=0.95\linewidth]{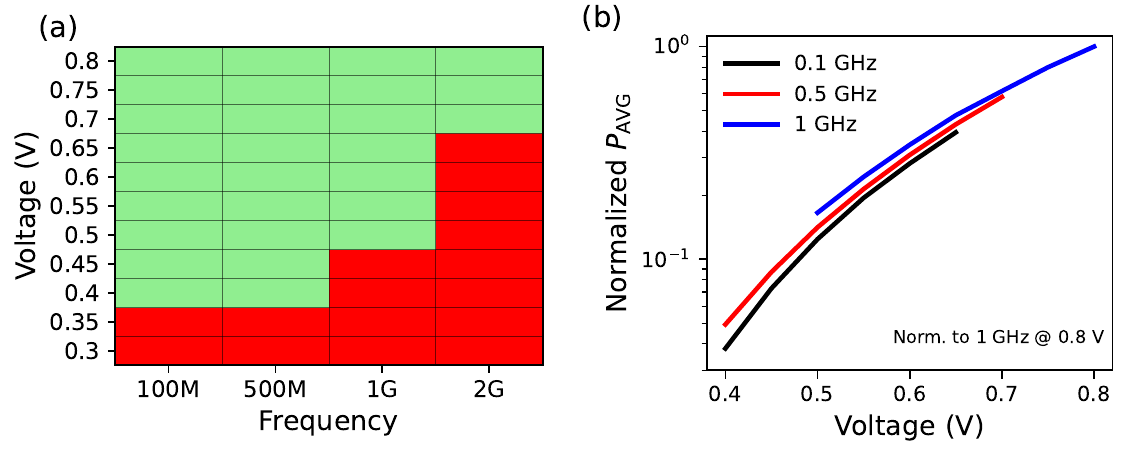}
    \caption{Voltage/frequency characterization of the AL systolic array with ideal four-phase PCKs. (a) Functional operating region. Green cells indicate correct operation, and red cells indicate failure. (b) Normalized average power of the functional points, normalized to the 1~GHz, 0.8~V case.}
    \label{fig:vf_region}
\end{figure}

Fig.~\ref{fig:vf_region}(a) shows that the minimum functional PCK amplitude increases with frequency. The AL systolic array operates correctly at 1~GHz with a PCK amplitude of 0.5~V, while 2~GHz requires approximately 0.7~V. Below these limits, the reduced device overdrive and signal swing prevent reliable charge recovery and logic evaluation within one clock period.

Fig.~\ref{fig:vf_region}(b) shows the corresponding normalized power trend. Only functionally correct points are used for energy comparison. At 1~GHz, reducing the PCK amplitude from 0.8~V to 0.5~V lowers the normalized average power from 1.0 to 0.165 while maintaining correct operation.

\subsection{Core Area and Energy-Area Analysis}

We estimate the MAC-core active area using a device-count model to avoid bias from synthesis mapping and optimization settings. The CMOS INV, BUF, AND, OR, and XOR gates use 2, 4, 6, 6, and 10 FinFET devices, respectively, while the corresponding AL gates use 6, 6, 8, 8, and 12 devices. Including the required AL phase-alignment buffers, the AL MAC core has an estimated 1.35$\times$ active-area overhead over the static CMOS MAC core.

Using the core-level power results in Table~\ref{AL_comparison}, the AL MAC core achieves a 42\% power reduction at 1~GHz, so its normalized energy per cycle is 0.58. We then compute the normalized core energy-area product (EAP) as
\[
\mathrm{Norm.~EAP}_{\mathrm{core}}
=
\frac{E_{\mathrm{AL,core}}}{E_{\mathrm{DIG,core}}}
\cdot
\frac{A_{\mathrm{AL,core}}}{A_{\mathrm{DIG,core}}}
=
0.78 .
\]
Thus, the proposed AL MAC core still achieves a 22\% lower EAP than the static CMOS MAC core after accounting for the estimated active-area overhead.

\section{CONCLUSIONS}
We present a complete adiabatic logic design methodology for MAC systolic arrays, from PFAL gate-level building blocks through system-level integration with digital peripherals. A novel resonant 4-phase power clock generator, co-designed with the AL load to eliminate external tank capacitors, drives the array at 1 GHz with over 60\% efficiency. 
In advanced FinFET nodes, picosecond-scale RC time-constants allow adiabatic circuits to maintain charge-recovery behavior at GHz frequencies, eliminating the historical speed limitation of adiabatic logic. 
On 16 nm FinFET technology, the AL MAC systolic array achieves 42\% and 36\% power reduction at the core and system levels, respectively. Even accounting for the 1.35$\times$ area overhead of dual-rail AL gates, the design achieves a 22\% lower energy-area product than its digital counterpart. Scaled to a 16$\times$16 array, the projected energy efficiency reaches 7.98 TOPS/W, surpassing state-of-the-art digital accelerators. Silicon implementation in advanced nodes is a natural next step.


\bibliographystyle{IEEEtran}
\bibliography{ref}

\end{document}